\documentclass[english,aps,preprint,nofootinbib,superscriptaddress]{revtex4-2}%
\usepackage[T1]{fontenc}
\usepackage[latin9]{inputenc}
\usepackage{amsmath}
\usepackage{amssymb}
\usepackage{babel}
\usepackage{amsfonts}
\usepackage{graphicx}
\usepackage{MnSymbol}

\begin{document}

\title{%Quantum field theory approach to relativistic wavepacket tunneling: a fully causal process
Relativistic Quantum Field Theory Approach to Wavepacket Tunneling: Lack of Superluminal Transmission}
 \author{M.\ Alkhateeb$^{*}$}
	\affiliation{Laboratoire de Physique Th\'eorique et Mod\'elisation, CNRS Unit\'e 8089, CY
		Cergy Paris Universit\'e, 95302 Cergy-Pontoise cedex, France}
		 \altaffiliation[Current address : ]{Research Unit Lasers and Spectroscopies (UR-LLS), naXys \& NISM, University of Namur, Rue de Bruxelles 61, B-5000 Namur, Belgium}
	\author{X. Gutierrez de la Cal}
	\affiliation{Departmento de Qu\'imica-F\'isica, Universidad del Pa\' is Vasco, UPV/EHU, 48940 Leioa, Spain}
		\affiliation{EHU Quantum Center, Universidad del Pa\' is Vasco, UPV/EHU, 48940 Leioa, Spain}
	\author{M. Pons}
	\affiliation{EHU Quantum Center, Universidad del Pa\' is Vasco, UPV/EHU, 48940 Leioa, Spain}
	\affiliation{Departamento de F\'isica Aplicada, Universidad del Pa\' is Vasco, UPV/EHU, 48013 Bilbao, Spain}
	\author{D. Sokolovski}
	\affiliation{Departmento de Qu\'imica-F\'isica, Universidad del Pa\' is Vasco, UPV/EHU, 48940 Leioa, Spain}
	\affiliation{EHU Quantum Center, Universidad del Pa\' is Vasco, UPV/EHU, 48940 Leioa, Spain}
	\affiliation{IKERBASQUE, Basque Foundation for Science, E-48011 Bilbao, Spain}
	\author{A.\ Matzkin}
	\affiliation{Laboratoire de Physique Th\'eorique et Mod\'elisation, CNRS Unit\'e 8089, CY
		Cergy Paris Universit\'e, 95302 Cergy-Pontoise cedex, France}

\begin{abstract}
		We investigate relativistic wavepacket dynamics for an electron tunneling
through a potential barrier employing space-time resolved solutions to
relativistic quantum field theory (QFT) equations. We prove by linking the QFT property of
micro-causality to the wavepacket behavior that the tunneling dynamics is fully causal, precluding
instantaneous or superluminal effects that have recently been reported in the
literature. We illustrate these results by performing numerical computations
for an electron tunneling through a potential barrier for standard tunneling
as well for Klein tunneling. In all cases (Klein tunneling \ or regular
tunneling across a standard or a supercritical potential) the transmitted
wavepacket remains in the causal envelope of the propagator, even when its
average position lies ahead of the average position of the corresponding
freely propagated wavepacket.

\end{abstract}
\maketitle

\newpage

%\email{mohammed.alkhateeb@cyu.fr}

%\email{alexandre.matzkin@cnrs.fr}

Tunneling is one of the most intriguing quantum phenomena. Although tunneling
underlies many important processes in about every area concerned by quantum
physics (see e.g. \cite{cm1,atto-r2,steinberg-prl,h2,atto-r1,ca2,cm2} for
recent observations), its precise mechanism has remained controversial
\cite{muga,recent-rev}. Despite experimental data coming from different areas,
from strong field tunneling ionization
\cite{keller2008,keller2012,atto-nosup-1,atto-r2,atto-r1} to cold atoms
\cite{steinberg-prl} neutron optics \cite{neutrons} or condensed matter
\cite{cmexp}, there seems to be no solution in view \cite{soko-recent} to the
tunneling time problem (the time spent by a particle inside the barrier), or
equivalently the arrival time (whether a particle that tunnels through a
barrier arrives earlier than a freely propagating particle). Indeed, due to
the ambiguity of measuring time in quantum mechanics -- there is no time
operator in the standard formalism -- any observed tunneling time will depend
on the model employed to extract the time interval from the observed data.

In particular, experiments involving electron photo-ionization have reported
results interpreted to indicate instantaneous tunneling times
\cite{keller2008,keller2012,atto-r2,atto-r1}. Such interpretations rely on
models that intrinsically involve disputed approximations
\cite{keitel-reconciling}, generally employing a non-relativistic and often
semiclassical framework. Perhaps somewhat more surprisingly, several works
based on a first-quantized relativistic framework
\cite{grobe-super,janner,winful,deleo,bernardini,gasparian,deleo2013,pollak2020,galapon}
have concluded on the possibility of superluminal arrival times for electrons.
Such superluminal transmissions could potentially bring serious issues with
causality, even though it is sometimes asserted that these effects do not seem
to lead to signaling \cite{pollak2020}. Other investigations carried out
within first quantized relativistic quantum mechanics have on the contrary not
noted any superluminal effects at the level of the wavefunction
\cite{korean,us-wp,math}.

In this work, we investigate the tunneling dynamics in a second quantized
framework. More specifically, we will employ a computational
relativistic\ quantum field theory (QFT) approach in order to follow the
space-time resolved dynamics of an electron tunneling through an electrostatic
potential barrier represented by a background field.\ The electron is modeled
as a wave-packet initially defined on a compact support launched towards a
potential barrier. We will prove that micro-causality of the fermionic quantum field
implies that the electron wavepacket density evolves causally,
thereby ensuring the absence of any superluminal
effects such as instantaneous tunneling times. The present method allows us to
treat on the same footing different types of tunneling effects: the familiar
one characterized by exponentially decaying waves inside the barrier, as well
as Klein tunneling (with undamped oscillating waves in the barrier) for
supercritical barriers (that is barriers with a potential above the
pair-production threshold).

Our approach is based on a computational QFT framework \cite{grobe-su-review},
recently extended to treat particle scattering across a finite barrier
\cite{AMpra22} (see also \cite{cqft-opt,lv} for related recent work). Since we
are interested in studying possibly superluminal phenomena, it makes sense
\cite{berry} to take initially ($t=0$) a wavepacket defined over a compact
support.\ As is well-known \cite{FV,fulling}, a relativistic quantum state defined on a compact
spatial support must contain both positive and negative energy components, so
that the initial state in Fock space is defined by%
\begin{equation}
\Vert\chi\rrangle=\int dp\big(g_{+}(p)b_{p}^{\dagger}+g_{-}(p)d_{p}^{\dagger
}\big)\Vert0\rrangle. \label{cs1}%
\end{equation}
Here $b_{p}^{\dagger}(t)$ and $d_{p}^{\dagger}(t)$ are the creation operators
for a particle and\ antiparticle of momentum $p$, and $b_{p}(t)$ (resp.
$d_{p}(t))$ are the corresponding annihilation operators; $\Vert0\rrangle$
defines the vacuum state, i.e. $b_{p}\Vert0\rrangle=d_{p}\Vert0\rrangle=0$,
and $g_{\pm}(p)$ are the expansion coefficients in momentum space. Since we
are dealing with a Dirac field the creation and annihilation operators
anti-commute, $[b_{p},b_{k}^{\dagger}]_{+}=[d_{p},d_{k}^{\dagger}]_{+}%
=\delta(p-k)$.
% Eq. (\ref{cs1}) implies there is initially a (small)
%probability to find a positron rather than an electron if the wavepacket is
%measured.\ However we can choose the wavepacket parameters so that for $t>0$
%only the electron part of $\Vert\chi\rrangle$ propagates towards the barrier
%(the positron part propagates in the opposite direction).

The particle density at any given time is given by the expectation value
\begin{equation}
\rho(t,x)=\llangle\chi\Vert\hat{\rho}(t,x)\Vert\chi\rrangle \label{tdd}%
\end{equation}
where the density operator $\hat{\rho}(t,x)$ is defined by
\begin{equation}
\hat{\rho}(t,x)=\hat{\Phi}^{\dagger}(t,x)\hat{\Phi}(t,x). \label{densdef}%
\end{equation}
$\hat{\Phi}(t,x)$ is the field operator suited to obtain the evolved compact
support Fock space state $\hat{\Phi}(t,x)=\int dpb_{p}(t)v_{p}(x)+\int
dpd_{p}(t)w_{p}(x).$ $v_{p}(x)$ and $w_{p}(x)$ are resp. the positive and
negative energy spinor eigenfunctions of the field-free Dirac Hamiltonian
[see Eq. (A-3) of Appdx A]. The free Dirac Hamiltonian
$H_{0}=-i\hbar c\alpha_{x}\partial_{x}+\beta mc^{2}$ has the corresponding eigenvalues
$\pm\left\vert E_{p}\right\vert =\pm\sqrt{p^{2}c^{2}+m^{2}c^{4}}$ ($\alpha$
and $\beta$ are the usual Dirac matrices \footnote{In one spatial dimension,
we can neglect spin-flip and replace $\alpha_{x}$ and $\beta$ by the Pauli
matrices $\sigma_{1}$ and $\sigma_{3}$ respectively.}, $m$ the electron mass
and $c$ the light velocity). The equal-time anti-commutators obey
\begin{equation}
\lbrack\hat{\Phi}^{\dagger}(t,x^{\prime}),\hat{\Phi}(t,x)]_{+}=\delta
(x^{\prime}-x) \label{et}%
\end{equation}
just like the familiar field operators of the free Dirac field \cite{greiner}
(see Appendix A for the proof of Eq. (\ref{et}) and the relation to the familiar QFT case). The
time-dependent creation and annihilation operators are obtained as prescribed
by our computational QFT\ method through \cite{grobe-su-review}%
	\begin{align}
	b_{p}(t)  &  =\int dk\left(  U_{v_{p}v_{k}}(t)b_{k}(0)+U_{v_{p}w_{k}}%
	(t)d_{k}^{\dagger}(0)\right) \label{bogu1}\\
	d_{p}^{\dagger}(t)  &  =\int dk\left(  U_{w_{p}v_{k}}(t)b_{k}(0)+U_{w_{p}%
		w_{k}}(t)d_{k}^{\dagger}(0)\right)  . \label{bogu2}%
\end{align}
$U$ is the unitary evolution operator of the full Hamiltonian $H=H_{0}+V(x).$
The barrier potential $V(x)$ is treated as background external field \cite{gitman}.\ We will
work for convenience with a rectangular-like model potential $V(x)=\frac{V_{0}}%
{2}\left[  \tanh(\left(  x+L/2\right)  /\epsilon)-\tanh(\left(  x-L/2\right)
/\epsilon)\right]  $ where $L$ is the barrier width and $\epsilon$ a
smoothness parameter. The unitary evolution operator elements, $U_{v_{k}w_{p}%
}(t)\equiv\left\langle v_{k}\right\vert \exp\left(  -iHt/\hbar\right)
\left\vert w_{p}\right\rangle $ are computed numerically on a discretized
space-time grid using a split operator \cite{split} method (the evolution
operator is split into a kinetic part propagated in momentum space and a
potential-dependent part solved in position space).

\begin{figure}[tb]
	\hspace*{-0.6cm}\includegraphics[scale=0.5]{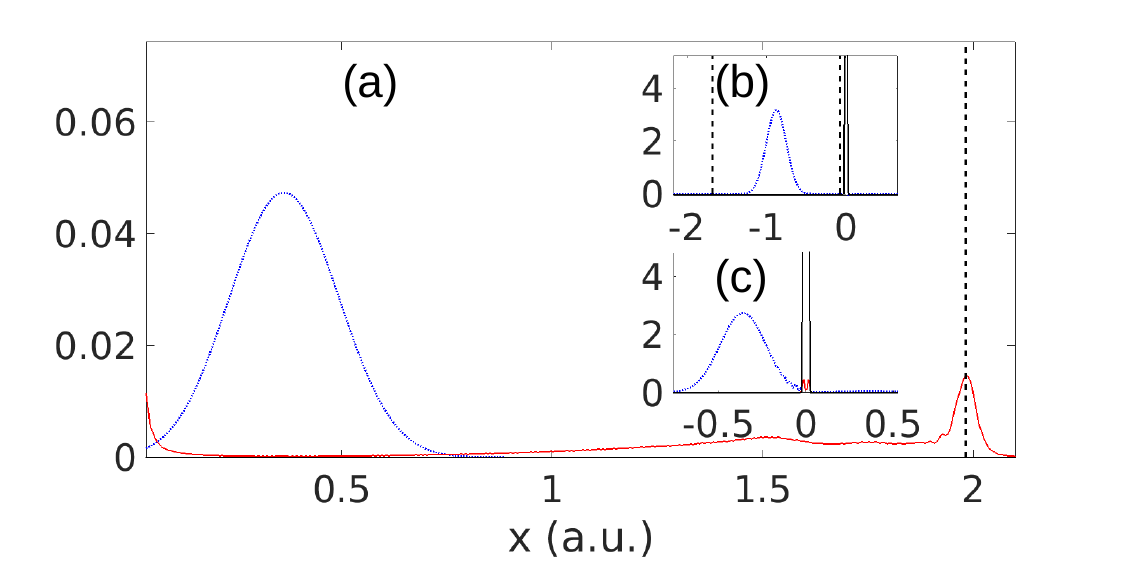}
\caption{(a) The density of the transmitted wavepacket is shown (dotted blue) as it is exiting the barrier ($t=1.5 \times10^{-2}$ a.u.) for a comparatively low potential ($V_{0} =0.5mc^{2}$) giving rise to standard tunneling, with a negligible pair creation rate (the electron density created by the potential is shown in red). The inset displays snapshots of the wavepacket dynamics at (b) $t=0$ and at (c) $t=1.5 \times10^{-2}$ a.u (note the transmitted wavepacket is hardly visible on that scale in (c)). The dotted vertical line in (b)
represents the right edge of the support $\mathcal{D}$ over which the initial wavepacket is defined. The same line in (a) and (c) represents the position of the light-cone emanating from this right edge at the time of the plot. The
initial wavepacket parameters in atomic units (a.u.) are $x_{0}= -120 \lambda$, $p_{0}=100$ a.u. and $ \mathcal {D}= 70 \lambda$ and for the barrier $L=4 \lambda$ and $\varepsilon= 0.3 \lambda$, where $\lambda=\hbar /mc$ is the Compton wavelength of the electron.}%
\label{fig-shallow}%
\end{figure}

An example of such a computation is shown in Fig.\ \ref{fig-shallow}. An
initial wave-packet given by the Dirac spinor $G(x)=\left(  \cos^{8}(\frac{x-x_{0}}{D})e^{ip_{0}%
x},0\right)  $ is defined to be non-zero only over the compact support
$x\in\mathcal{D}$ where $\mathcal{D}=[x_{0}-D\pi/2,x_{0}+D\pi/2]$ is localized
to the left of the barrier, with an initial mean momentum such that the
electron wave-packet moves towards the right as time evolves. By projecting
this spatial profile over the free Dirac basis $v_{p}(x)$ and $w_{p}(x)$ we obtain
the coefficients $g_{\pm}(p)$ of Eq. (\ref{cs1}) needed to define the initial
second quantized wave-packet\footnote{Recall that the first quantized
wavepacket is obtained from the Fock space state through $\chi
(t,x)=\llangle0\Vert\hat{\Phi}(t,x)\Vert\chi\rrangle$ \cite{schweber,compact}} which is in turn fed in Eq. (\ref{tdd}) in order to obtain
the space-time resolved density $\rho(t,x)$. $\rho(t,x)$ can be parsed in
several ways (see Appendix B). In particular for distances sufficiently far
from the barrier, the density represents the sum of the transmitted or
reflected electron wavepacket and the electron density due to pair production
(which is asymptotically small for barriers of height $V_{0}\ll2mc^{2}$).

\begin{figure}[t]
	\hspace*{-0.6cm}\includegraphics[scale=0.5]{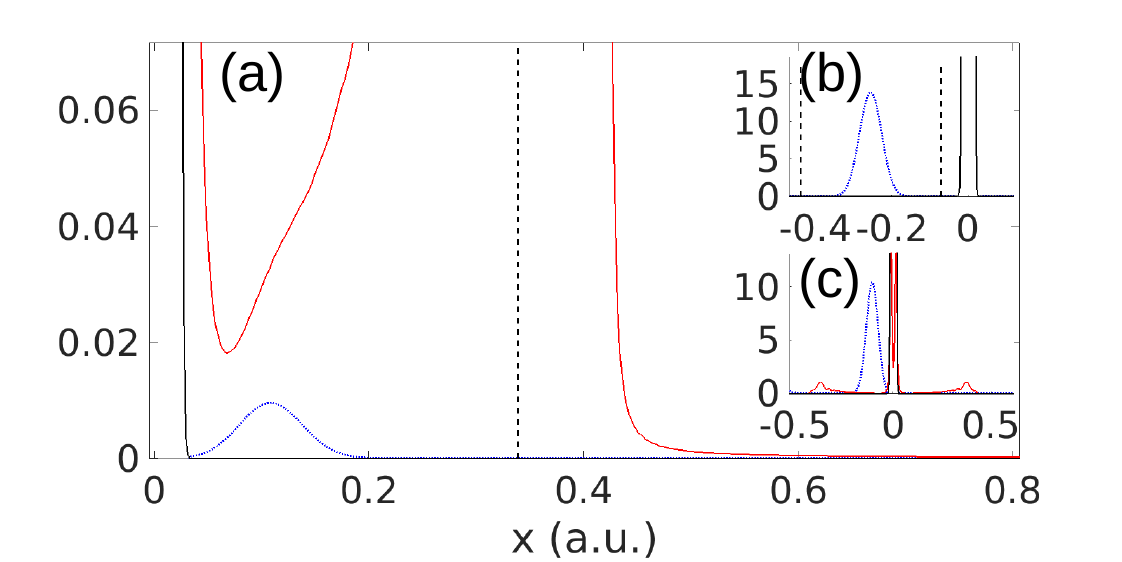}\caption{Similar to Fig.
\ref{fig-shallow} but for a stronger potential $V_{0}=1.77mc^{2}$. (a) The
density of the transmitted wavepacket (dotted blue) is overshadowed by the
electron density due to pair creation and appears as a bump in the overall
particle density (shown in red). The inset displays snapshots of the
wavepacket dynamics at (b) $t=0$ and at (c) $t=3 \times 10^{-3}$ (the time of
the plot (a)). The initial wavepacket parameters are (in a.u.)
$x_{0}=-35 \lambda$, $p_{0}=200$  and $\mathcal{D}=16 \lambda$ and for the barrier $L=4 \lambda$ and $\varepsilon= 0.3 \lambda$, where $\lambda=\hbar / mc$. }%
\label{fig-regular}%
\end{figure}

Fig. \ref{fig-shallow} shows the transmitted wavepacket as well as the
electron density due to pair production for a comparatively ``low'' potential ($V_{0}=0.5mc^{2}$). Snapshots of the density
evolution are given in the inset; leaving aside pair production, this
situation is a QFT\ account of the familiar tunneling dynamics, where most of
the incoming electron amplitude is reflected and only a very small amplitude
is transmitted.\ Fig.\ \ref{fig-regular} shows the situation for a higher
barrier ($V_{0}=1.77mc^{2}$) at $t_{p}=3\times10^{-3}$ a.u.:\ pair-production
is still small (the total number of electrons due to pair production is
$N_{vac}(t_{p})/2=0.31$ [see Eq. (A-33) of Appdx B]), but the transmitted
wavepacket amplitude is even smaller and overshadowed by the electron density produced by the
barrier and appears as a bump in the overall electron density.
%Recall that
%potential barriers below the $2mc^{2}$ threshold have a very small pair
%production probability that is nevertheless comparable to the typical
%amplitude of a transmitted wavepacket.
Note that some of the works \cite{grobe-super,janner,winful,deleo,bernardini,gasparian,deleo2013,pollak2020,galapon} investigating
relativistic tunneling within the first quantized approximation have computed
numerical results for barrier heights in cases in which QFT\ calculations show that the tiny
amplitude of the transmitted wavepacket is completely obscured by the larger
(or much larger if supercritical barriers are considered) electron density
produced by the barrier.

Fig. \ref{fig-regular} also shows the light cone, emanating from the right
edge ($x=x_{0}+D\pi/2$) of the initial wavepacket density distribution; it can
be seen that although \ the electron is in the relativistic regime (the mean
velocity of the initial distribution is $0.83c$), the transmitted wavepacket
remains well inside the light cone. This is an illustration of a very general
result hinging on micro-causality of relativistic quantum fields: observables that are
space-like separated commute.\ If $O(t,x)$ and $O^{\prime}(t^{\prime
},x^{\prime})$ are two obervables, $[\hat{O}^{\prime},\hat{O}]=0\;\;$%
for$\;\;c^{2}\left(  t^{\prime}-t\right)  ^{2}-\left(  x^{\prime}-x\right)
^{2}<0$, thereby implying that observations made at space-like separated points are
independent. It is straightforward to verify that micro-causality
holds here: by recalling that a general observable is built from a bilinear form of field operators
\cite{greiner,pad-book}, the commutator
$[\hat{O}^{\prime},\hat{O}]$ for arbitrary observables can be seen to be proportional to the field anti-commutator
$[\hat{\Phi}^{\dagger}(t^{\prime},x^{\prime}),\hat{\Phi}(t,x)]_{+}$ (a particular
instance is given in Eq. (\ref{microrho}) of Appdx A for the important case of density observables).
These
field anti-commutators vanish for space-like separated points: this
can be seen by Lorentz-boosting (to another reference frame for which
$t^{\prime}\neq t$) the equal-time anti-commutator of Eq.
(\ref{et}). As proved in Appendix A, this anti-commutator vanishes for space-like separated events for the free case and also in the presence of background fields.

\begin{figure}[b]
		\hspace*{-0.6cm}\includegraphics[scale=0.5]{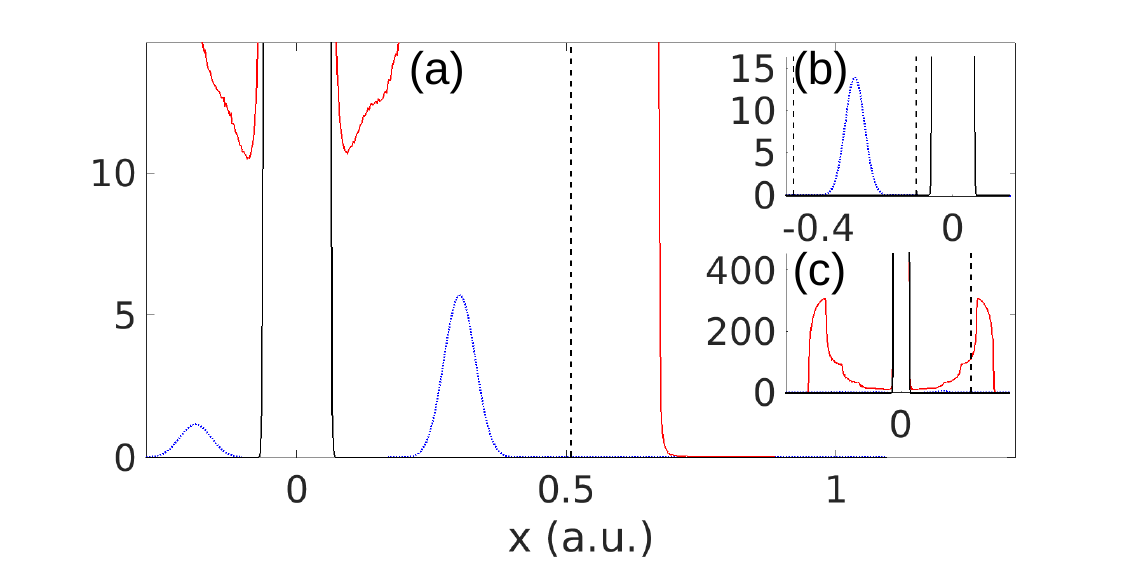}
	\caption{Similar to Figs.
		\ref{fig-shallow} and \ref{fig-regular} but for a potential $V_{0}=9 mc^{2}$ above the supercritical limit, giving rise to Klein tunneling. (a) The electron wavepacket density  is shown (dotted blue) at $t= 4.5 \times 10^{-3} $ a.u. well after the transmitted wavepacket (centered at $x\approx 0.3$ a.u.) has exited the barrier (solid vertical lines). Note that the transmitted wavepacket density is significantly larger than the one of the reflected wavepacket (centered at $x=-0.19$ a.u. and moving toward the left). (b) The initial wavepacket (light blue) is shown along with the support $\mathcal{D}$ (dashed lines) and the barrier. (c) The plot (a) is zoomed out in order to visualize the electron density due to pair production (red line). The wavepacket is not visible at this scale. The
		initial wavepacket parameters in a.u. are $x_{0}=-40 \lambda$,$p_{0}%
		=450$ a.u. and $ \mathcal{D}=16 \lambda$  and for the barrier $L=16 \lambda$ and $\varepsilon=0.3 \lambda$ with $\lambda= \hbar / mc$. }%
	\label{fig-klein}%
\end{figure}

	We can now show that micro-causality imposes the causality of the tunneling dynamics in the
following way. Let us choose $\hat{O}$ as an intervention on
the wavepacket density at $t=0,$ $\hat{O(0,\mathcal{D})}=\int_{\mathcal{D}}dx\hat{\Phi}^{\dagger
}(0,x)f(x)\hat{\Phi}(0,x)$ where $f(x)$ is an arbitrary positive-definite real function that
modifies the profile of the initial wavepacket. $O$ is now defined over 
the domain $\mathcal{D}$ rather than at a single point $x$. Note that $f(x)$ must be
chosen such that $\llangle\chi\Vert\hat{O}\Vert\chi\rrangle=\int
_{\mathcal{D}}dxf(x)\chi^{\dagger}(x)\chi(x)$ remains normalized. We will set
$\hat{O}^{\prime}=\hat{\rho}(t^{\prime},x^{\prime})$ to be the density [Eq.
(\ref{densdef})] at a space-like separated point $(x^{\prime}>c\left(
t^{\prime}-t\right)  +x_{0}+D\pi/2$) to the right of the barrier and
sufficiently far from it $\left(  x^{\prime}\gg L/2\right) $. Note that
$x_{0}+D\pi/2$ being the right edge of $\mathcal{D}$, 
any point of the initial wavepacket density is space-like separated form $(t^{\prime},x^{\prime})$.
%It is straightforward to show that $[\hat{O}^{\prime},\hat{O}]=0$ holds for such observables.
By relying on micro-causality, it can be established that
\begin{equation}
	\llangle\chi\Vert\hat{O}^{\prime}(t^{\prime},x^{\prime})\hat{O}(0,\mathcal{D})\Vert
	\chi\rrangle=\llangle0\Vert\hat{O}^{\prime}(t^{\prime},x^{\prime})\Vert0\rrangle.
	\label{nos}%
\end{equation}
This means that the electron density at $(t^{\prime},x^{\prime})$ is given by
a vacuum expectation value and does not depend in any way on the initial wavepacket
density or any operation one would perform on the wavepacket at $t=0$ (the vacuum density is non-vanishing due to the electrons produced by the barrier). These results are proved in Appendix C.

Eq. (\ref{nos}) linking the causal behavior of the tunneled wavepacket to micro-causality is our main theoretical result and implies that tunneling
cannot be superluminal nor instantaneous once relativistic QFT constraints are
taken into account. It is noteworthy that this result does not depend on the
shape, width or height of the background potential (see Appendix C) -- it also holds in
particular for more complicated potentials than the smooth rectangular barrier
we have employed here. This result holds of course for all types of tunneling
-- for regular tunneling, as in Figs. \ref{fig-shallow} or \ref{fig-regular}, or
for\ Klein tunneling.

Klein tunneling takes place for supercritical potentials
($V_{0}>2mc^{2}$) and wavepacket energies for which $\left(  E-V_{0}\right)
^{2}>m^{2}c^{4}$; in this case the transmission of the electron wavepacket is
mediated by pair production \cite{KP-grobe-su,AMpra22} giving rise to an oscillating
density inside the barrier.\ These modulations in pair-production give rise to
a transmitted wavepacket with an undamped amplitude (as opposed to an
exponentially decreasing tranmission in the case of regular tunneling).
Relative to the freely propagated wavepacket, the transmitted Klein tunneled
one can be accelerated by the barrier (since the negative
energy wavepacket components
see a potential well \cite{calogeracos}) but never faster than light, since
our result Eq. (\ref{nos}) holds for any type of potential barrier.\ A computation
illustrating Klein tunneling is given in Fig. \ref{fig-klein}, for $V_{0}=9mc^{2}$.

\begin{figure}[t]
\includegraphics[scale=0.15]{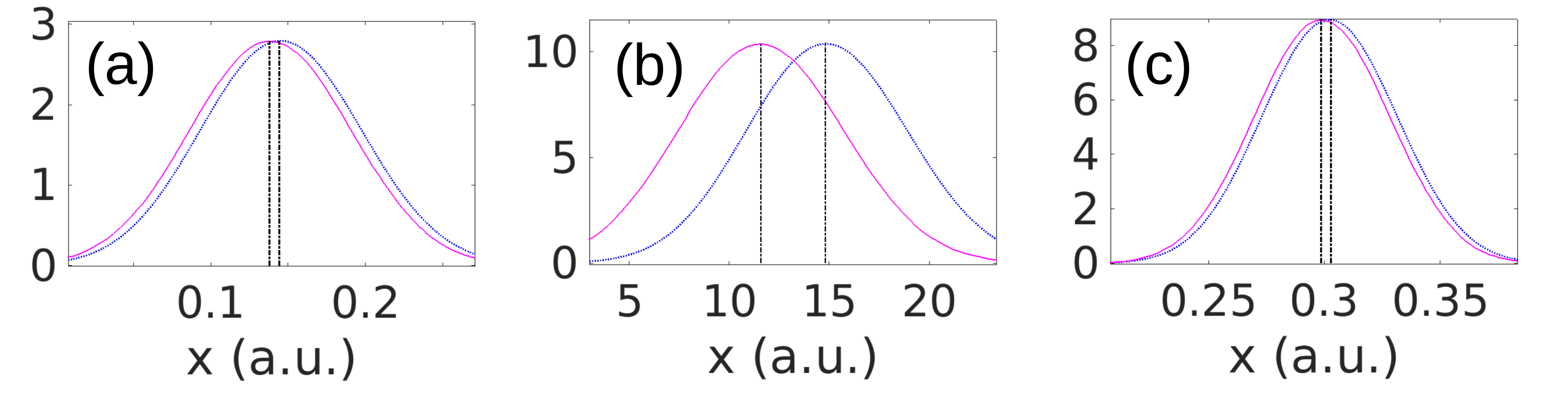}
	\caption{(a), (b) and (c)
		display for each case considered respectively in Figs. \ref{fig-shallow},
		\ref{fig-regular} and \ref{fig-klein}, the position of the transmitted peak
		along with the position of the same initial wavepacket that would have evolved
		freely. The vertical dotted lines indicate the averages $\left\langle
		X_{\mathrm{tr}}(t)\right\rangle $ and $\left\langle X_{\mathrm{free}%
		}(t)\right\rangle $  (see text for details).}%
	\label{fig-vsfree}%
\end{figure}

Finally, since it is often stated that tunneling can be superluminal and we
have shown here that this is contrary to the predictions obtained from a
space-time resolved relativistic QFT approach to spin-1/2 fermions, it is
worthwhile briefly recalling on which gounds such assertions have been made.
We must first discard models based on non-relativistic frameworks, like the
Schr\"odinger equation, for which propagation is indeed instantaneous
\cite{hegerfeldt}, or semi-classical approximations to it. Experimental
results, in particular those involving the attoclock technique in strong field
ionization (see e.g. \cite{keller2008,keller2012,atto-r2,atto-r1}), have
usually relied on such models when estimating tunneling times. Second, there
is no unambiguous manner to define a tunneling time\ \cite{soko-recent} and
the various quantities that have been proposed (phase delays, dwell times,
Larmor times, time operators) lead to conflicting results and may by
construction yield superluminal values, including when they are employed with
relativistic wave equations
\cite{grobe-super,winful,deleo,bernardini,deleo2013,galapon}.

Third, some first quantized works based on relativistic wave equations have
suggested \cite{janner,pollak2020} superluminal transmission based on the fact
that the transmitted wavepacket arrives on average earlier than the freely
propagating one, i.e.\ $\left\langle X_{\mathrm{tr}}(t)\right\rangle
>\left\langle X_{\mathrm{free}}(t)\right\rangle $ where in $\left\langle
X_{\mathrm{tr}}(t)\right\rangle $ the average is taken over the transmitted
peak only (it is a conditional expectation value). Asserting that the
transmitted wavepacket travels faster on this basis is only possible if one
associates a conditional wavepacket (the transmitted one) with a single
particle. While reasoning in this manner might be disputed even from within a
standard quantum mechanics perspective (it is far from obvious that a fraction
of a wavefunction can be associated with a single particle), it is clearly not
compatible with a QFT based framework. According to QFT, a particle at each
space-time point of a wavepacket is seen as a field excitation at that
particular point, and the field excitation at that point is causally related
to the field excitation at some other space-time point, in particular to the
field excitation at a different position in a given reference frame. In the
three numerical examples given here  we also have $\left\langle X_{\mathrm{tr}}(t)\right\rangle
>\left\langle X_{\mathrm{free}}(t)\right\rangle $ (see Fig. \ref{fig-vsfree})
while still being constrained by Eq.\ (\ref{nos}).

To sum up, we have investigated the tunneling wavepacket dynamics for an
electron within a relativistic QFT framework in which the
barrier is modeled as a background field. We have shown that if the electron
wavepacket is initially ($t=0$) localized to the left of the barrier, the
electron density at a space-like separated point to the right of the barrier
does not depend on the presence or absence of the wavepacket at $t=0,$ thereby
precluding any superluminal effects related to tunneling. We have numerically
computed the space-time resolved electron density in typical cases of
tunneling with potentials below, close to or above the supercritical value. We
hope our results will contribute in clarifying the models and approximations
employed when accounting for results involving traversal or detection times in tunneling related effects.

\bigskip

\textit{Acknowledgments.}
This project has received funding from the European Union's Horizon 2020 research and innovation programme under the Marie Sk{\l}odowska-Curie grant agreement No 101034383. We are grateful for grant PID2021-126273NB-I00, funded by MCIN/AEI/10.13039/ 501100011033 and  ``ERDF A way of making Europe''. We acknowledge financial support from the Basque Government, grant No. IT1470-22. MP acknowledges support from the Spanish Agencia Estatal de Investigacion, grant No. PID2022-141283NB-100.

\onecolumngrid
\newpage

\appendix
\renewcommand{\theequation}{A-\arabic{equation}}
\setcounter{equation}{0}

\section*{Appendix A - Field operators and equal-time anti-commutators}

\label{AppA} The field operator $\hat{\Phi}(t,x)$ is given in terms of the
annihilation operators of particles and antiparticles by:%
\begin{equation}
	\hat{\Phi}(t,x)=\int dp\left(  \hat{b}_{p}(t)v_{p}(x)+\hat{d}_{p}%
	(t)w_{p}(x)\right)  \label{phi}%
\end{equation}
and its Hermitian conjugate applied to the dual Fock states is given by:
\begin{equation}
	\hat{\Phi}^{\dagger}(t,x)=\int dp\left(  \hat{b}_{p}^{\dagger}(t)v_{p}%
	^{\dagger}(x)+\hat{d}_{p}^{\dagger}(t)w_{p}^{\dagger}(x)\right)  ,
	\label{phidag}%
\end{equation}
where $v_{p}(x)$ and $w_{p}(x)$ are the solutions of the free Dirac equation
in one spatial dimension given by%
\begin{equation}%
	\begin{split}
		v_{p}(x)  &  =%
		\begin{pmatrix}
			1\\
			\frac{cp}{mc^{2}+E_{p}}%
		\end{pmatrix}
		e^{ipx}\\
		w_{p}(x)  &  =%
		\begin{pmatrix}
			1\\
			\frac{cp}{mc^{2}-E_{p}}%
		\end{pmatrix}
		e^{-ipx}%
	\end{split}
	\label{fonctions}
\end{equation}

In the \emph{field free }case, the time evolution of the creation and
annihilation operators is trivial ($\hat{b}_{p}(t)=e^{iE_{p}t}\hat{b}_{p}$,
$\hat{d}_{p}^{\dagger}(t)=e^{-iE_{p^{\prime}}t}\hat{d}_{p}^{\dagger}$, etc.)
\ and the equal-time anti-commutator reads%
\begin{equation}%
	\begin{split}
		&  \left[  \hat{\Phi}^{\dagger}(x),\hat{\Phi}(y)\right]  _{+}=\\
		&  \left[  \int dp\hat{b}_{p}^{\dagger}v_{p}^{\dagger}(x)e^{iE_{p}t}+\int
		dp\hat{d}_{p}^{\dagger}w_{p}^{\dagger}(x)e^{-iE_{p}t},\int dp^{\prime}\hat
		{b}_{p^{\prime}}^{}v_{p^{\prime}}(y)e^{-iE_{p^{\prime}}t}+\int
		dp\hat{d}_{p^{\prime}}^{}w_{p^{\prime}}(y)e^{-iE_{p^{\prime}}t}\right]
		_{+}%
	\end{split}
	\label{nusf}%
	\end{equation}
Using the anti-commutation relations
\begin{equation}%
	\begin{split}
		\lbrack\hat{b}_{p}^{\dagger},\hat{b}_{p^{\prime}}]_{+}  &  =[\hat{d}%
		_{p}^{\dagger},\hat{d}_{p^{\prime}}]_{+}=\delta(p-p^{\prime}),\\
		\lbrack\hat{b}_{p}^{\dagger},\hat{d}_{p^{\prime}}]_{+}  &  =[\hat{d}%
		_{p}^{\dagger},\hat{b}_{p^{\prime}}]_{+}=\delta(p-p^{\prime}),
	\end{split}
	\label{creanncomm}%
\end{equation}
and
\begin{equation}%
	\begin{split}
		v_{p}^{\dagger}(x)v_{p}(y)  &  =e^{ip(y-x)}\\
		w_{p}^{\dagger}(x)w_{p}(y)  &  =e^{-ip(x-y)},
	\end{split}
\end{equation}
we obtain
\begin{equation}
	\left[  \hat{\Phi}^{\dagger}(x),\hat{\Phi}(y)\right]  _{+}=\int dp(e^{ip(y-x)}%
	+e^{ip(x-y)})
\end{equation}
which leads to Eq. (\ref{et}).

In the presence of a \emph{background potential}, the equal-time
anti-commutation relation
\begin{equation}
	\left[  \hat{\Phi}^{\dagger}(t,x),\hat{\Phi}(t,y)\right]  _{+}=\left[  \int
	dp\left(  \hat{b}_{p}^{\dagger}(t)v_{p}^{\dagger}(x)+\hat{d}_{p}^{\dagger
	}(t)w_{p}(x)^{\dagger}\right)  ,\int dp\left(  \hat{b}_{p}(t)v_{p}(x)+\hat
	{d}_{p}(t)w_{p}(x)\right)  \right]  _{+} \label{commut-d}%
\end{equation}
invoves the anti-commutators of the type
\begin{equation}%
	\begin{split}
		&  [\hat{b}_{p_{1}}^{\dagger}(t),b_{p_{2}}(t)]=\\
		&  \left[  \int dp_{1}^{\prime}\left(  U_{v_{p_{1}}w_{p_{1}^{\prime}}}^{\ast
		}\hat{b}_{p_{1}^{\prime}}^{\dagger}+U_{v_{p_{1}}w_{p_{1}^{\prime}}}^{\ast}%
		\hat{d}_{p_{1}^{\prime}}\right)  ,\int dp_{2}^{\prime}\left(  U_{v_{p_{2}%
			}v_{p_{2}^{\prime}}}\hat{b}_{p_{2}^{\prime}}+U_{v_{p_{2}}w_{p_{2}^{\prime}}%
		}\hat{d}_{p_{2}^{\prime}}^{\dagger}\right)  \right]  _{+}.
	\end{split}
\end{equation}
Using Eq. (\ref{bogu1}), one obtains%
\begin{equation}%
	\begin{split}
		&  \left[  \hat{b}_{p_{1}}^{\dagger}(t),b_{p_{2}}(t)\right]  _{+}\\
		&  =\int d_{p_{1}^{\prime}}\left(  U_{v_{p_{1}}v_{p_{1}^{\prime}}}^{\ast
		}U_{v_{p_{2}}v_{p_{1}^{\prime}}}+U_{v_{p_{1}}w_{p_{1}^{\prime}}}^{\ast
		}U_{v_{p_{2}}w_{p_{1}^{\prime}}}\right) \\
		&  =\int d_{p_{1}^{\prime}}\left(  \langle v_{p_{2}}|\hat{U}|v_{p_{1}^{\prime
		}}\rangle\langle v_{p_{1}^{\prime}}|\hat{U}^{\dagger}|v_{p_{1}}\rangle+\langle
		v_{p_{2}}|\hat{U}|w_{p_{1}^{\prime}}\rangle\langle w_{p_{1}^{\prime}}|\hat
		{U}^{\dagger}|v_{p_{1}}\rangle\right) \\
		&  =\langle v_{p_{2}}|\hat{U}\hat{U}^{\dagger}|v_{p_{1}}\rangle=\langle
		v_{p_{2}}|v_{p_{1}}\rangle=\delta(p_{1}-p_{2}),
	\end{split}
	\label{bs}%
\end{equation}
where in the last line, we used the completeness relation:%
\[
\int dp^{\prime}\left(  |v_{p^{\prime}}\rangle\langle v_{p^{\prime}%
}|+|w_{p^{\prime}}\rangle\langle w_{p^{\prime}}|\right)  =1
\]
and the orthonormality of the solutions of the free Dirac equation. Similarly,
we find that%
\begin{equation}
	\left[  \hat{d}_{p_{1}}^{\dagger}(t),d_{p_{2}}(t)\right]  _{+}=\delta
	(p_{1}-p_{2}). \label{ds}%
\end{equation}
Pluging-in these anti-commutators into Eq. (\ref{commut-d}) leads to%
\begin{equation}
	\left[  \hat{\Phi}^{\dagger}(t,x),\hat{\Phi}(t,y)\right]  _{+}=\int
	dp(e^{ip(y-x)}+e^{ip(x-y)}) \label{nus}%
\end{equation}
and hence again to Eq. (\ref{et}).

We now compute the equal-time \textit{commutator} for the density observables $\left[  \hat{\rho}(t,x),\hat{\rho}(t,y)\right]$ where
$\hat{\rho}(t,x)=\hat{\Phi}^{\dagger}(t,x)\hat{\Phi}(t,x)$. This commutator  can be written in
terms of the field anti-commutators as:%
%\[%
\begin{equation}
\begin{split}
	\left[  \hat{\rho}(t,x),\hat{\rho}(t,y)\right]   &  =\hat{\Phi}^{\dagger
	}(t,x)\left(  \left[  \hat{\Phi}(t,x),\hat{\Phi}^{\dagger}(t,y)\right]
	\hat{\Phi}(t,y)+\hat{\Phi}^{\dagger}(t,y)\left[  \hat{\Phi(t,x)},\hat
	{\Phi}(t,y)\right]  \right)  \\
	&  +\left(  \left[  \hat{\Phi}^{\dagger}(t,x),\hat{\Phi}^{\dagger
	}(t,y)\right] \hat{\Phi}(t,y)+\hat{\Phi}^{\dagger}(t,y)\left[  \hat{\Phi
	}^{\dagger}(t,x),\hat{\Phi}(t,y)\right]  \right)  \hat{\Phi}(t,x)\\
	&  =\hat{\Phi}^{\dagger}(t,x)\left[  \hat{\Phi}(t,x),\hat{\Phi}^{\dagger
	}(t,y)\right]  _{+}\Phi(t,y)-\hat{\Phi}^{\dagger}(t,y)\left[  \hat{\Phi
	}^{\dagger}(t,x),\hat{\Phi}(t,y)\right]  _{+}\hat{\Phi}(t,x),
\end{split}
\label{microrho}
\end{equation}
%\]
Since the equal time anti-commutator, given by Eq. (\ref{nus}), vanishes for for $x\neq y$, we obtain an equal-time commutator for
the number density operator that also vanishes for $x\neq y$.%
%\begin{equation}
%\left[  \hat{\rho}(t,x),\hat{\rho}(t,y)\right]=0.
%\label{microrho2}
%\end{equation}

Note that the anti-commutators (\ref{nusf}) or (\ref{nus}) also hold for the
standard field operators $\hat{\Psi}(t,x)$ and its Hermitian conjugate
$\hat{\Psi}^{\dagger}(t,x)$; in the field free case,
\begin{equation}
	\left[  \hat{\Psi}^{\dagger}(t,x),\hat{\Psi}(t,y)\right]  _{+}=\delta(x-y)
	\label{normalc}%
\end{equation}
is derived in many textbooks (e.g. \cite{greiner,pad-book}) as an instance of
the unequal-time anti-commutator obtained in terms of propagators. However the
proof used here to obtain Eq. (\ref{nus}) also works similarly to obtain Eq.
(\ref{normalc}) by recalling that \ \cite{grobe-su-review}
\begin{equation}%
	\begin{split}
		\hat{\Psi}(t,x)  &  =\int dp\hat{b}_{p}(t)v_{p}(x)+\int dp\hat{d}_{p}%
		^{\dagger}(t)w_{p}(x)\equiv\hat{\Psi}_{pa}(t,x)+\hat{\Psi}_{an}^{\dagger
		}(t,x)\\
		\hat{\Psi}^{\dagger}(t,x)  &  =\int dp\hat{b}_{p}^{\dagger}(t)v_{p}^{\dagger
		}(x)+\int dp\hat{d}_{p}(t)w_{p}^{\dagger}(x)\equiv\hat{\Psi}_{pa}^{\dagger
		}(t,x)+\hat{\Psi}_{an}(t,x)
	\end{split}
	\label{stfo}%
\end{equation}
where $\hat{\Psi}_{{pa}}$ and $\hat{\Psi}_{an}$ are the positive frequency
parts of $\hat{\Psi}$ and $\hat{\Psi}^{\dagger}$ respectively (linked to
particle and anti-particle annihilation). In terms of these operators, we can
write $\Phi$ and $\Phi^{\dagger}$ as \cite{compact}
\begin{equation}%
	\begin{split}
		\hat{\Phi}(t,x)  &  =\hat{\Psi}_{pa}(t,x)+\left(  \hat{\Psi}_{an}(t,x)\right)
		^{\ast T}\\
		\hat{\Phi}^{\dagger}(t,x)  &  =\hat{\Psi}_{pa}^{\dagger}(t,x)+\left(
		\hat{\Psi}_{an}^{\dagger}(t,x)\right)  ^{\ast T}.
	\end{split}
\end{equation}

\section*{Appendix B - Derivation of the density expression}

\label{AppB}

We derive here the expression of the particle density, given by Eq.
(\ref{tdd}). The density is the expectation value of the density operator [Eq.
(\ref{densdef})] when the initial Fock space state is the wavepacket
$\Vert\chi\rrangle=\int dp\big(g_{+}(p)\hat{b}_{p}^{\dagger}+g_{-}(p)\hat
{d}_{p}^{\dagger}\big)\Vert0\rrangle$. We therefore write%
\begin{equation}%
	\begin{split}
		\rho(t,x)  &  =\llangle\chi\Vert\hat{\rho}(t,x)\Vert\chi\rrangle\\
		&  =\llangle0\Vert\int dp\big(g_{+}^{\ast}(p)\hat{b}_{p}+g_{-}^{\ast}%
		(p)\hat{d}_{p}\big)\hat{\rho}(t,x)\int dp\big(g_{+}(p)\hat{b}_{p}^{\dagger
		}+g_{-}(p)\hat{d}_{p}^{\dagger}\big)\Vert0\rrangle.
	\end{split}
\end{equation}
and insert the expressions of $\hat{\Phi}(t,x)$ and $\hat{\Phi}^{\dagger
}(t,x)$ given in Eqs. (\ref{phi})-(\ref{phidag}), yielding%

\begin{align}
	\rho(t,x)=  &  \llangle0\Vert\int dp\big(g_{+}^{\ast}(p)\hat{b}_{p}%
	+g_{-}^{\ast}(p)\hat{d}_{p}\big)\\
	&  \Big\{\iint dp_{1}dp_{2}v_{p_{1}}^{\dagger}(x)v_{p_{2}}(x)\hat{b}_{p_{1}%
	}^{\dagger}(t)\hat{b}_{p_{2}}(t)\label{zr1}\\
	+  &  \iint dp_{1}dp_{2}w_{p_{1}}^{\dagger}(x)w_{p_{2}}(x)\hat{d}_{p_{1}%
	}^{\dagger}(t)\hat{d}_{p_{2}}(t)+\label{zr3}\\
	&  \Big(\iint dp_{1}dp_{2}v_{p_{1}}^{\dagger}(x)w_{p_{2}}(x)\hat{b}%
	_{p}^{\dagger}(t)\hat{d}_{p}(t)+HC\Big)\Big\}\label{zr4}\\
	&  \int dp\big(g_{+}(p)\hat{b}_{p}^{\dagger}+g_{-}(p)\hat{d}_{p}^{\dagger
	}\big)\Vert0\rrangle.
\end{align}

This density can be parsed as a sum of three terms, each term corresponding to
the expectation value obtained for each line, Eqs. (\ref{zr1})-(\ref{zr3}).
particle density, $\rho_{1}(t,x)$, antiparticle density, $\rho_{2}(t,x)$and a
"mixed term", $\rho_{3}(t,x)$:
\begin{equation}
	\rho(t,x)=\rho_{1}(t,x)+\rho_{2}(t,x)+\rho_{3}(t,x). \label{rhotot}%
\end{equation}

Let us first compute the expectation value of the operator written in Eq.
(\ref{zr1}).\ Using Eq. (\ref{bogu1}), we obtain
\begin{equation}%
	\begin{split}
		\rho_{1}(t,x)=\llangle0\Vert &  \int dp\big(g_{+}^{\ast}(p)\hat{b}_{p}%
		+g_{-}^{\ast}(p)\hat{d}_{p}\big)\Big\{\iint dp_{1}dp_{2}v_{p_{1}}^{\dagger
		}(x)v_{p_{2}}(x)\int dp^{\prime}\Big(U_{v_{p_{1}}v_{p^{\prime}}}^{\ast}%
		(t)\hat{b}_{p^{\prime}}^{\dagger}+U_{v_{p_{1}}w_{p^{\prime}}}^{\ast}(t)\hat
		{d}_{p^{\prime}}\Big)\\
		&  \int dp^{\prime}\Big(U_{v_{p_{2}}v_{p^{\prime}}}(t)\hat{b}_{p^{\prime}%
		}+U_{v_{p_{2}}w_{p^{\prime}}}(t)\hat{d}_{p^{\prime}}^{\dagger}\Big)\Big\}\int
		dp\big(g_{+}(p)\hat{b}_{p}^{\dagger}+g_{-}(p)\hat{d}_{p}^{\dagger}%
		\big)\Vert0\rrangle
	\end{split}
\end{equation}
which expands to
\begin{equation}%
	\begin{split}
		\rho_{1}(t,x)  &  =\llangle0\Vert\int\cdots\int dq_{1}dq_{1}^{\prime}%
		dq_{2}dq_{2}^{\prime}dp_{1}dp_{2}g_{-}^{\ast}(q_{1})g_{-}(q_{2})U_{v_{p_{1}%
			}w_{q_{1}^{\prime}}}^{\ast}(t)U_{v_{p_{2}}w_{q_{2}^{\prime}}}(t)v_{p_{1}%
		}^{\dagger}(x)v_{p_{2}}(x)\hat{d}_{q_{1}}\hat{d}_{q_{1}^{\prime}}\hat
		{d}_{q_{2}^{\prime}}^{\dagger}\hat{d}_{q_{2}}^{\dagger}\Vert0\rrangle\\
		&  +\llangle0\Vert\int\cdots\int dq_{1}dq_{1}^{\prime}dq_{2}dq_{2}^{\prime
		}dp_{1}dp_{2}g_{+}^{\ast}(q_{1})g_{+}(q_{2})U_{v_{p_{1}}w_{q_{1}^{\prime}}%
		}^{\ast}(t)U_{v_{p_{2}}w_{q_{2}^{\prime}}}(t)v_{p_{1}}^{\dagger}(x)v_{p_{2}%
		}(x)\hat{b}_{q_{1}}\hat{d}_{q_{1}^{\prime}}\hat{d}_{q_{2}^{\prime}}^{\dagger
		}\hat{b}_{q_{2}}^{\dagger}\Vert0\rrangle\\
		&  +\llangle0\Vert\int\cdots\int dq_{1}dq_{1}^{\prime}dq_{2}dq_{2}^{\prime
		}dp_{1}dp_{2}g_{+}^{\ast}(q_{1})g_{+}(q_{2})U_{v_{p_{1}}v_{q_{1}^{\prime}}%
		}^{\ast}(t)U_{v_{p_{2}}v_{q_{2}^{\prime}}}(t)v_{p_{1}}^{\dagger}(x)v_{p_{2}%
		}(x)\hat{b}_{q_{1}}\hat{b}_{q_{1}^{\prime}}^{\dagger}\hat{b}_{q_{2}^{\prime}%
		}\hat{b}_{q_{2}}^{\dagger}\Vert0\rrangle.
	\end{split}
\end{equation}
Using the anti-commutation relations of creation and annihilation operators
\begin{equation}%
	\begin{split}
		&  \llangle0\Vert\hat{d}_{q_{1}}\hat{d}_{q_{1}^{\prime}}\hat{d}_{q_{2}%
			^{\prime}}^{\dagger}\hat{d}_{q_{2}}^{\dagger}\Vert0\rrangle=\delta
		_{q_{1}^{\prime}q_{2}^{\prime}}\delta_{q_{1}q_{2}}-\delta_{q_{1}q_{2}^{\prime
		}}\delta_{q_{1}^{\prime}q_{2}}\\
		&  \llangle0\Vert\hat{b}_{q_{1}}\hat{d}_{q_{1}^{\prime}}\hat{d}_{q_{2}%
			^{\prime}}^{\dagger}\hat{b}_{q_{2}}^{\dagger}\Vert0\rrangle=\delta_{q_{1}%
			q_{2}}\delta_{q_{1}^{\prime}q_{2}^{\prime}}\\
		&  \llangle0\Vert\hat{b}_{q_{1}}\hat{b}_{q_{1}^{\prime}}^{\dagger}\hat
		{b}_{q_{2}^{\prime}}\hat{b}_{q_{2}}^{\dagger}\Vert0\rrangle=\delta_{q_{1}%
			q_{2}^{\prime}}\delta_{q_{2}q_{2}^{\prime}},
	\end{split}
\end{equation}
we get%
\begin{equation}%
	\begin{split}
		\rho_{1}(t,x)  &  =\int dq\big|g_{-}(q)\big|^{2}\int dq\left(  \int
		U_{v_{p}w_{q}}(t)v_{p}(x)\right)  ^{\dagger}\left(  \int U_{v_{p}w_{q}%
		}(t)v_{p}(x)\right) \\
		&  +\int dq\big|g_{+}(q)\big|^{2}\int dq\left(  \int U_{v_{p}w_{q}}%
		(t)v_{p}(x)\right)  ^{\dagger}\left(  \int U_{v_{p}w_{q}}(t)v_{p}(x)\right) \\
		&  +\left(  \int dpdqg_{+}(p)U_{v_{p}v_{q}}v_{p}(x)\right)  ^{\dagger}\left(
		\int dpdqg_{+}(p)U_{v_{p}v_{q}}v_{p}(x)\right) \\
		&  -\left(  \int dpdqg_{-}(p)U_{v_{p}w_{q}}v_{p}(x)\right)  ^{\dagger}\left(
		\int dpdqg_{-}(p)U_{v_{p}w_{q}}v_{p}(x))\right)
	\end{split}
\end{equation}
Using the normalization of the initial QFT state yields%
\begin{equation}%
	\begin{split}
		\rho_{1}(t,x)  &  =\int dq\left(  \int U_{v_{p}w_{q}}(t)v_{p}(x)\right)
		^{\dagger}\left(  \int U_{v_{p}w_{q}}(t)v_{p}(x)\right) \\
		&  +\left(  \int dpdqg_{+}(p)U_{v_{p}v_{q}}(t)v_{p}(x)\right)  ^{\dagger
		}\left(  \int dpdqg_{+}(p)U_{v_{p}v_{q}}(t)v_{p}(x)\right) \\
		&  -\left(  \int dpdqg_{-}(p)U_{v_{p}w_{q}}(t)v_{p}(x)\right)  ^{\dagger
		}\left(  \int dpdqg_{-}(p)U_{v_{p}w_{q}}(t)v_{p}(x))\right)  .
	\end{split}
	\label{rho1}%
\end{equation}

The first line in the expression of $\rho_{1}(t,x)$ represents the electron
density created by the background potential due to the vacuum excitation while
the second line represents the density corresponding to the incoming particle.
The third line represents the modulation in the number density of the created
particles due to the incident particle wave packet. The terms $\rho_{2}(t,x)$
and $\rho_{3}(t,x)$ are computed similarly, yielding%
\begin{equation}%
	\begin{split}
		\rho_{2}(t,x)  &  =\int dp\left(  \int dqU_{w_{p}v_{q}}(t)w_{p}(x)\right)
		^{\dagger}\left(  \int dqU_{w_{p}v_{q}}(t)w_{p}(x)\right) \\
		&  +\left(  \int dpdqg_{-}(q)U_{w_{p}w_{q}}(t)w_{p}(x)\right)  ^{\dagger
		}\left(  \int dpdqg_{-}(q)U_{w_{p}w_{q}}(t)w_{p}(x)\right) \\
		&  -\left(  \int dpdqg_{+}(q)U_{w_{q}w_{p}}(t)w_{p}(x)\right)  ^{\dagger
		}\left(  \int dpdqg_{+}(q)U_{w_{q}v_{q}}(t)w_{p}(x)\right)
	\end{split}
	\label{rho2}%
\end{equation}
and%
\begin{equation}%
	\begin{split}
		\rho_{3}(t,x)  &  =2\Re\Big(\int dpdqg_{-}^{\ast}(q)U_{w_{p}w_{q}}^{\ast
		}(t)g_{+}(q)U_{v_{p}v_{q}}w_{q}^{\dagger}(x)v_{p}(x)\Big)\\
		&  +2\Re\Big(\int dpdqg_{-}^{\ast}(q)U_{w_{p}v_{q}}^{\ast}(t)g_{+}%
		(q)U_{v_{p}w_{q}}w_{q}^{\dagger}(x)v_{p}(x)\Big)
	\end{split}
	\label{rho3}%
\end{equation}
$\rho_{2}(t,x)$ is the counterpart of $\rho_{1}(t,x)$ for the positron density
while $\rho_{3}(t,x)$ involves cross terms between positive and negative
energy modes of the initial wave-packet. $\rho_{3}(t,x)$ cancels the infinite
tails of $\rho_{1}(t,x)$ and $\rho_{2}(t,x)$. When integrated over the entire
space however the contribution of this term vanishes, ensuring that $\rho$
obeys%
\begin{equation}
	\int dx\rho(t,x)=\int dx\rho_{1}(t,x)+\int dx\rho_{2}(t,x)
\end{equation}
which is the sum of the particle and antiparticle numbers.

Note that in the expressions of $\rho_{1}$ and $\rho_{2},$ there is only a
single term that does not depend on the wavepacket (the first line in Eqs.
(\ref{rho1}) and (\ref{rho2})).\ Hence by subsuming these two lines into
$\rho_{vac}(t,x)$, the total density can also be parsed as
\begin{equation}
	\rho(t,x)=\rho_{vac}(t,x)+\rho_{wp}(t,x).
\end{equation}
By removing $\rho_{vac}(t,x)$ from the computed total density one can thus
visualize the wavepacket contribution to the density. The total number of
particles $N(t)$, obtained by integrating the density over all space, can be
also be parsed as%
\begin{equation}
	N(t)=\int dx\rho(t,x)=N_{vac}(t)+1\label{numdef}%
\end{equation}
where the wavepacket counts as one particle. \ $N(t)$ can also be written as
the normal-ordered expectation value of the number operator $\hat{N}(t)$
written in the standard form%
\begin{equation}
	\hat{N}(t)=\int dp\left(  \hat{b}_{p}^{\dagger}(t)\hat{b}_{p}(t)+d_{p}%
	^{\dagger}(t)d_{p}(t)\right)  .
\end{equation}

\section*{Appendix C - Causality condition on the wavepacket}

\label{AppC}

First, let us choose the observable $\hat{O}(t=0)$ as an observable that modifies the initial wavepacket
on the compact support $\mathcal{D}$ over which the wavepacket is defined. Let us set
\begin{equation}
\hat{O}(t=0,\mathcal{D})=\int_{\mathcal{D}}dx\hat{\Phi}^{\dagger}(0,x)f(x)\hat{\Phi}(0,x),
\end{equation}
where $f(x)$ is an arbitrary function defined on  $\mathcal{D}$ that modifies the profile of the initial
wavepacket while conserving the initial norm.\ $f(x)$ reshapes the initial
wavepacket (e.g. , $f(x)=\sqrt{2}\theta(x-x_{0})$ keeps only the right half of
the wavepacket while increasing its amplitude so as to preserve
normalisation). Indeed, the expectation value of $\hat{O}$ is computed as%
\begin{equation}
	\llangle\chi\Vert\hat{O}(0)\Vert\chi\rrangle=\int dxf(x)\chi^{\dagger
	}(0,x)\chi(0,x)=1. \label{zzz}%
\end{equation}
This can be seen by starting from the initial QFT state given by Eq.
(\ref{cs1}) and
\begin{equation}
	\hat{\Phi}(0,x)=\int dp\left(  \hat{b}_{p}(0)v_{p}(x)+\hat{d}_{p}%
	(0)w_{p}(x)\right)  .
\end{equation}
One then gets, using $\hat{b}_{p}\hat{b}_{p^{\prime}}^{\dagger}\Vert
0\rrangle=\delta(p-p^{\prime})\Vert0\rrangle$ and $\hat{d}_{p}\hat
{d}_{p^{\prime}}^{\dagger}\Vert0\rrangle=\delta(p-p^{\prime})\Vert0\rrangle,$%
\begin{equation}
	\hat{\Phi}(0,x)\Vert\chi\rrangle=\int dp\left(  g_{+}(p)v_{p}(x)+g_{-}%
	(p)w_{p}(x)\right)  \Vert0\rrangle=\chi(0,x)\Vert0\rrangle; \label{phichi2}%
\end{equation}
similarly we have
\begin{equation}
	\llangle\chi\Vert\hat{\Phi}^{\dagger}(0,x)=\llangle0\Vert\chi^{\dagger}(0,x)
	\label{phichi3}%
\end{equation}
from which Eq. (\ref{zzz}) follows.

Second, let $\hat{O^{\prime}}(t^{\prime},x^{\prime})$ be the density operator at the
spacetime point $(t^{\prime},x^{\prime})$,%
\begin{equation}
	\hat{O^{\prime}}(t^{\prime},x^{\prime})=\hat{\Phi}^{\dagger}(t^{\prime
	},x^{\prime})\hat{\Phi}(t^{\prime},x^{\prime}).
\end{equation}
$x^\prime$ is chosen to lie far to the right of the barrier, and such that
$(t^{\prime},x^{\prime})$ is pace-like separated from $\mathcal{D}$ at $t=0$.
The left hand side of Eq. (\ref{nos}) can be written as
\begin{equation}
	\llangle\chi\Vert\hat{O^{\prime}}(t^{\prime},x^{\prime})\hat{O}(t=0)\Vert
	\chi\rrangle=\llangle\chi\Vert\hat{O^{\prime}}(t^{\prime},x^{\prime}%
	)\int_{\mathcal{D}}f(x)\hat{\Phi}^{\dagger}(0,x)\hat{\Phi}(0,x)\Vert
	\chi\rrangle. \label{lhs}%
\end{equation}
Since both $\hat{\Phi}^{\dagger}(t^{\prime},x^{\prime})$ and $\hat{\Phi
}(t^{\prime},x^{\prime})$ anti-commute with $\hat{\Phi}(0,x)$ given that the
two spacetime points $(0,x)$ and $(t^{\prime},x^{\prime})$ are space-like, we
have%
\begin{equation}
	\llangle\chi\Vert\hat{O^{\prime}}(t^{\prime},x^{\prime})\hat{O}\Vert
	\chi\rrangle=\llangle\chi\Vert\int_{\mathcal{D}}f(x)\hat{\Phi}^{\dagger
	}(0,x)\hat{O^{\prime}}(t^{\prime},x^{\prime})\hat{\Phi}(0,x)\Vert\chi\rrangle.
	\label{lhs2}%
\end{equation}
Inserting Eqs. (\ref{phichi2})-(\ref{phichi3}) in Eq. (\ref{lhs2}), one
obtains%
\begin{equation}
	\llangle\chi\Vert\hat{O^{\prime}}(t^{\prime},x^{\prime})\hat{O}\Vert
	\chi\rrangle=\llangle0\Vert\hat{O^{\prime}}(t^{\prime},x^{\prime})\int
	_{D}dxf(x)\chi^{\dagger}(0,x)\chi(0,x)\Vert0\rrangle.
\end{equation}
Eq. (\ref{nos}) is recovered by using Eq. (\ref{zzz}). Note that a similar
proof can be obtained for observables built from bilinear forms of the
standard field operators $\hat{\Psi}(t,x)$ and $\hat{\Psi}^{\dagger}(t,x)$
defined in Eq. (\ref{stfo}).

\end{document}